\documentclass[epsf,useAMS,usenatbib]{mn2e}
\usepackage{epsf}
\usepackage{amsmath}
\usepackage{graphicx}
\newcommand{\simgt}{\lower.5ex\hbox{$\; \buildrel > \over \sim \;$}}
\newcommand{\simlt}{\lower.5ex\hbox{$\; \buildrel < \over \sim \;$}}

\newcommand{\baredth}{\;\overline{\raise1.0pt\hbox{$'$}\hskip-6pt
\partial}\;}
\newcommand{\edth}{\;\raise1.0pt\hbox{$'$}\hskip-6pt\partial\;}

\begin{document}
\title{Validity of strong lensing statistics for constraints on the galaxy evolution model}
\author[Akiko Matsumoto and Toshifumi Futamase]{$^1$Akiko Matsumoto and $^2$Toshifumi Futamase\thanks{E-mail:
$^1$a-matsu@astr.tohoku.ac.jp:
$^2$tof@astr.tohoku.ac.jp} 
\\
Astronomical Institute, Tohoku University, Sendai 980-8578, Japan.\\
}

\pagerange{\pageref{firstpage}--\pageref{lastpage}} \pubyear{2007}

\maketitle

\label{firstpage}

\begin{abstract}
\ \ We examine the usefulness of the strong lensing statistics to   
constrain the evolution of the number density of lensing galaxies by adopting the values of the cosmological parameters 
determined by recent WMAP observation.  
For this purpose, we employ the lens-redshift test proposed by
 Kochanek (1992) and constrain the parameters in two evolution models, 
simple power-law model characterized by the power law 
indexes $\nu_{n}$ and $\nu_{v}$ and the evolution model by Mitchell et al. (2005) 
based on CDM structure formation scenario.
We use the well-defined lens sample from the Sloan Digital Sky Survey
 (SDSS) and this is similarly sized samples used in the previous
 studies.
Furthermore, we adopt the velocity dispersion function of early-type galaxies 
based on SDSS DR1 and DR5.
It turns out that the indexes of power-law model are consistent with the
 previous studies, thus our results indicate the mild evolution in the number and velocity
 dispersion of early-type galaxies out to $z=1$.
However we found that the values for p and q used by Mitchell et al. are
 inconsistent with the presently available observational data. More
 complete sample is necessary to withdraw more realistic determination        
 on these parameters.

\end{abstract}



\section{INTRODUCTION}\label{(1)}

\ Since the pioneering work by Turner, Ostriker and Gott (Turner, Ostriker
\& Gott 1984), the strong lensing statistics is known as a powerful tool 
to constrain the cosmological parameters, especially the cosmological constant
(Fukugita, Futamase, Kasai \& Turner 1992), as well as to investigate 
the evolution of number density of lensing galaxies
(\cite{b13}, \cite{b14}, \cite{b15}, \cite{b4},
\cite{b16}, and \cite{b17}).
However this method is not competitive with
other approaches such as Cosmic Microwave Background(CMB). In fact 
recent observation by WMAP(e. g. \cite{b52}) determined the cosmological parameters 
within the accuracy of a few percent. 
The reason is mainly because of the shortage of observed lens systems as well as 
the partial knowledge of the redshift information of the sources and 
lensing galaxies. 

Specifically, the current largest complete lens sample comes from 
the Cosmic Lens All Sky Survey (CLASS; \cite{b23}, \cite{b24}), 
an extension of the earlier Jodrell Bank/Very Large Array Astrometric Survey
(JVAS; \cite{b20}, \cite{b21} and \cite{b22}), but the number of lensing events are 
only 13 lensed radio sources. On the other hand, the SDSS Quasar Lens Search (SQLS) 
has already discovered approximately 20 new strongly lensed quasars, 
so this sample is expected to become the largest statistical sample of 
strongly lensed quasars.

On the other hand, there have been no well motivated and established theoretical 
models for the evolution of the number density of the lensing galaxies, 
and there seems no definite 
observational conclusion if the number density evolves or not. 

Thus it will be timely to investigates the evolution of the number density 
of the lensing galaxies by adopting the values for the cosmological parameters 
determined by WMAP and using a new sample based on SQLS. 

In this paper, we focus on two typical redshift evolution models: 
One model is described by the power-law evolution for the number density and the characteristic
velocity dispersion as $(1+z)^{\nu_{n}}$ and $(1+z)^{\nu_{v}}$,
respectively. (see Chae \& Mao (2003) in detail).
The other is the model based on CDM structure formation scenario 
used by Mitchell et al. (2005) which we call Mitchell's model.  
We use `lens-redshift test` proposed by Kochanek (see Kochanek (1992)) to calculate 
the likelihood as a function of the parameters in the models. 
This method has advantage not to use the total lensing probability and thus 
not to require the knowledge of the magnification bias in the sample. 
We also use the SQLS data whose separation is under 2 arcsec as the lens
sample in order to select only the events caused by a single galaxy which 
is known to be well modeled by Singular Isothermal Sphere(SIS) or 
Singular Isothermal Ellipsoid(SIE). We have used the parameters in the velocity 
dispersion function given by SDSS DR1 and DR5. 
Thus we avoid ambiguities of the lens model as 
possible as we can.

The organization of this paper is as follows. 
In section 2, we explain some basic of lens statistics 
and show the lens models and the evolution models we adopt.
In section 3, we describe the data that we use.
We calculate maximum likelihood value and confidence level of the parameters 
of each evolution model using the differential lensing probability.
In section 4, we present the results.
Finally some discussion is given in section 5. 
Throughout this paper we adopt $\Lambda$ dominant totally flat cosmology
with the matter density $\Omega_{m}=0.27$.

\section{BASIC OF STRONG LENSING STATISTICS}

\subsection{The optical depth}

\ \ Following to Turner, Ostriker \& Gott (1984), we consider the light ray propagating 
from source (QSO) to observer. 
The differential probability $d\tau$ of the ray encountering a lens in traveling the path of the redshift interval 
$dz_{L}$ is given by
\begin{equation}\label{eq:dt}
d\tau=n_{L}(z_{L})\sigma_{L}\frac{cdt}{dz_{L}}dz_{L}B
\end{equation} 
where $n_{L}(z_{L})$ is number density of the lens (see 2.3),
$\sigma_{L}$is the lensing cross section (see \ref{(2.2)}), and B is the
magnification bias (see \cite{b1}).
The quantity $\frac{cdt}{dz_{L}}$ is calculated in the FLRW geometry to be
\begin{equation}
\frac{cdt}{dz_{L}}
=\frac{R_{0}}{1+z_{L}}\frac{1}{\sqrt{\mathstrut \Omega _{m}(1+z_{L})^{3}+\Omega_{\Lambda} }}
\end{equation}
where $R_0=1/H_{0}$ is Hubble distance, $\Omega_{m}$ is total mass density of 
the universe, $\Omega_{\Lambda }\equiv \Lambda c^{2}/3H_{0}^{2}$ is the normalized 
cosmological constant, and $z_{L}$ is lens redshift.

Using (\ref{eq:dt}), the differential optical depth of lensing in 
traversing $dz_{L}$ with angular separation between $\Delta\theta$ 
and $\Delta\theta+d\Delta\theta$ is calculated as $d^{2}\tau/dz_{L}d\Delta\theta$.

\subsection{The lens models}\label{(2.2)}
\ \ We use a singular isothermal sphere (SIS) and a singular isothermal
ellipsoid (SIE) (\cite{b2}) as  our lens model. \\
\ \ The SIS model is widely used, so we don't explain its property. (see,
for example, \cite{b1} and \cite{b30}).\\
\ \ The SIE model can be solved analytically (Asada et al. 2003).
We refer the readers to Asada et al. (2003) for specific calculation.
We only show the cross sections of this model for  4images ($\sigma_{L4im}$) and 2images ($\sigma_{L2im}$) here.
\begin{equation}
\sigma_{L4im}=\frac{3\pi}{8}\frac{4e^{2}}{\sqrt{1-e^{2}}}\theta_{E}^{2}
\end{equation}
\begin{equation}
\sigma_{L2im}=\pi\sqrt{1-e^{2}}\theta_{E}^{2}-\sigma_{L4im}
\end{equation}
where $e$ represents ellipticity and this ellipticity is constrained as $0\leq e <1/5$.
 The SIS model is recovered in the $e=0$ for $\sigma_{L2im}$ case  
and this upper constrain $e<1/5$ comes from the fact that the density
 contours must be convex, which is reasonable for an isolated relaxed system.
And $\theta_{E}$ is Einstein radius for SIS lens and can be written as
\begin{equation}
\theta_{E} = 4\pi\Bigl(\frac{\sigma}{c}\Bigr)^{2}\frac{D_{LS}}{D_{OS}}
\end{equation} 
where $\sigma$ is one-dimensional velocity dispersion 
and $D_{LS}$ and $D_{OS}$ are the angular diameter distance between the
lens and the source, the observer and the source, respectively.

\subsection{Number density of galaxy lens}
  The number density of galaxies with velocity dispersion (we call this
function velocity function after this) lying between $\sigma$ and
$\sigma+d\sigma$ is assumed to be described by modified Schechter
function (\cite{b3}, \cite{b4}) of the form,
\begin{equation}
\label{eq:velocity function}
\phi(\sigma)d\sigma=\phi_{\ast}\Bigl(\frac{\sigma}{\sigma_{\ast}}\Bigr)^{\alpha}\exp\Bigl[-\Bigl(\frac{\sigma}{\sigma_{\ast}}\Bigr)^{\beta}\Bigr]\frac{\beta}{\Gamma(\alpha/\beta)}\frac{d\sigma}{\sigma}
\end{equation}
where $\phi_{\ast}$ is the integrated number density of galaxies,
$\sigma_{\ast}$ is the characteristic velocity dispersion, $\alpha$ is the low-velocity 
power-low index and $\beta$ is the high-velocity exponential cut-off index.
These parameters in the velocity function, $\phi_{\ast}, \alpha, \beta$
and $\sigma_{\ast}$, are determined by observations.
$\Gamma(x)$ represents the gamma function.
The present day comoving number density of galaxies can be calculated by 
integrating (\ref{eq:velocity function}).

For these parameters, we use the measured velocity dispersion function 
from SDSS DR1 (\cite{b3}; \cite{b4}) and DR5 (\cite{b5}) for early-type galaxies.
The measured values that Chae (2007) outlined are as follows, 
\begin{eqnarray}
(\phi_{\ast}, \sigma_{\ast}, \alpha, \beta)_{DR1}=&&[(4.1\pm 0.3)\times
 10^{-3}\ h^{3}\ \mathrm{Mpc^{-3}}, \nonumber \\
&&88.8 \pm  17.7\ \mathrm{km}\ \mathrm{s^{-1}}, \nonumber \\
&&6.5\pm 1.0,\ 1.93\pm 0.22]
\label{eq:DR1}
\end{eqnarray}
\begin{eqnarray}
(\phi_{\ast}, \sigma_{\ast}, \alpha, \beta)_{DR5}=&&[8.0 \times 10^{-3}
 h^{3}\ \mathrm{Mpc^{-3}}, \nonumber \\
&&161 \pm  5\ \mathrm{km}\ \mathrm{s^{-1}}, \nonumber \\
&&2.32\pm 0.10,\ 2.67\pm 0.07]
\label{eq:DR5}
\end{eqnarray}
where $h$ is the Hubble constant in units of 100 $\mathrm{km/s}$ $\mathrm{Mpc^{-3}}$.

\subsection{The lens redshift distribution}\label{(2.3)}
 In this subsection we show two evolution models of number density of galaxies 
for (\ref{eq:velocity function}).

\subsubsection{The power-law evolution model}
 The first evolution model is a power-law type as
most authors adopted (for example, \cite{b6} ).
In this model, we allow the number density of galaxies $\phi_{\ast}$ and 
the velocity dispersion $\sigma_{\ast}$ in (\ref{eq:velocity function}) to 
vary with redshift as 
\begin{equation}
\phi_{\ast}(z)=\phi_{\ast}(1+z)^{\nu_{n}}
\end{equation}
\begin{equation}
\sigma_{\ast}(z)=\sigma_{\ast}(1+z)^{\nu_{v}}
\end{equation}
The no evolution model corresponds to $(\nu_{n}=0,\ \nu_{v}=0)$.
Note that the case ($\nu_{n}>0$, $\nu_{v}<0$) is consistent with the hierarchical 
model of galaxy formation with
bottom-up assembly of structure.

\subsubsection{Mitchell's evolution model}
 The next evolution model is  proposed by Mitchell et al. (2005).
We briefly explain this model.
They have taken into consideration the number density evolutions by combining 
a halo mass function with a velocity dispersion as a function of mass and redshift.
They use the prediction by the extended Press-Schechter model of structure formation, 
which is calibrated by N-body simulations (\cite{b8}) as the halo mass function.
This halo mass function at epoch z is introduced by Sheth \& Tormen (1999),
\begin{equation}
n(M;\ z)=\frac{\bar \rho}{M}\frac{d \ln \nu }{dM}A(p)[1+(q\nu )^{-p}]
\biggl(\frac{q\nu }{2\pi}\biggl)^{1/2}\exp\biggl(\frac{-q\nu }{2}\biggr)
\label{eq:st-func}
\end{equation}
where $\bar\rho$ is the mean density, $\nu (z)=\delta _{c}^{2}/\sigma_{\delta}^{2}(M,z)$, 
$\delta_{c}=1.686$ is the extrapolated linear overdensity of a spherical top-hat 
perturbation at the time it collapse, and $\sigma_{\delta}^{2}(M,z)$ is 
the variance of the density field at epoch z in linear perturbation theory, 
smoothed with a top hat filter of radius $R(M) = (3M/4\pi\bar\rho)^{1/3}$. 
The parameters, $(p,\ q,\ A(p))$, are given by N-body simulations
in cold dark matter models as $p=0.3,\ q=0.75,\ A(p)=0.3222$ (see \cite{b8}).

Next, we show the velocity dispersion (\cite{b9}; \cite{b10}) used in
Mitchell et al. (2005),
\begin{equation}
\sigma(M,z)=92.3\bigtriangleup
 _{vir}(z_{f})^{1/6}E(z_{f})^{1/3}\biggl(\frac{M}{10^{13}h^{-1}M_{\odot}}\biggr)^{1/3}  \mathrm{km}\ \mathrm{s^{-1}}
\label{eq:sigma}
\end{equation}
where 
\begin{equation}
\bigtriangleup_{vir}(z)=18\pi^{2}+82[\Omega(z)-1]-39[\Omega(z)-1]^{2},
\end{equation}
and
\begin{equation}
\Omega(z)=\frac{(1+z)^{3}}{E^{2}(z)}.
\end{equation}
where $E(z)=H(z)/H_{0}=[\Omega_{m}(1+z)^{3}+\Omega_{\Lambda}]^{1/2}$ and
$z_{f}$ is the mean formation redshift for a halo mass M observed at
redshift z (see \cite{b4} and \cite{b12} ).
Combining equation (\ref{eq:st-func}) and (\ref{eq:sigma}) yields the velocity function at redshift function.

Following above mentioned procedure, we can compute the ratio of the velocity dispersion function at two epochs,
$\phi(\sigma;z)/\phi(\sigma;0)$, as a function of cosmological parameters. 
This ratio can be combined with the velocity function  (\ref{eq:velocity function}).

\begin{table*}
\centering
\begin{tabular}{llllllll}
 \hline
Number & Name          & $z_{s}$ & $z_{l}$ & $\Delta\theta$     & Grade & $N_{im}$ &  References \\
(1)    & (2)           & (3)          & (4)            & (5)     & (6)     & (7)     & (8)\\
\hline
1      & SDSS0806+2006 & $1.540$  & $0.573$  & $1.40$ & A     & 2        & 1\\
2      & SBS0909+523   & $1.38$  & $0.83$  & $1.17$ & A     & 2    & 2\\
3      & SDSS0924+0219 & $1.52$  & $0.39$  & $1.75$ & A     & 4   & 3, 4\\
4     & FBQ0951+2635  & $1.24$  & $0.20$  & $1.11$ & A     & 2     & 5, 6\\
5     & B1030+074     & $1.54$  & $0.60$  & $1.65$ & A     & 2      & 7\\
6     & B1152+200     & $1.02$  & $0.44$  & $1.59$ & A     & 2     & 8\\
7     & SDSS1226-0006 & $1.12$  & $0.52$  & $1.26$ & A     & 2     & 9\\
8      & SDSS1332+0347   & $1.445$  & $0.191$  & $1.14$ & A     & 2        & 10\\
9      & SDSS1353+1138   & $1.629$  & $0.3$  & $1.41$ & A     & 2        & 11\\
10      & SDSS1406+6126 & $2.13$  & $0.27$  & $1.98$ & A     & 4        & 12\\
11      & SBS1520+530   & $1.86$  & $0.72$  & $1.59$ & A     & 2        & 13\\
12     & HST15433+5352 & $2.09$  & $0.50$  & $1.18$ & A     & 2R    & 14\\
13     & MG1549+3047   & $1.17$  & $0.11$  & $1.7$ & A     & R       &
			     15, 16\\
\hline
\end{tabular}
\caption{The columns are: (1) lens number; (2) lens name; (3) source
 redshift, $z_{s}$; (4) lens redshift, $z_{l}$; (5) lens separation, $\Delta\theta$, in arcseconds; (6) grade for the likelihood that the object is a lens: A=I'd bet my life, B=I'd bet your life, and C=I'd bet your life and you should worry (CASTLES); (7) Number of images corresponding to each source component, E means extended and R means there is an Einstein ring (CASTLES); (8) references.
\newline
List of references: \newline $1$ - Inada
et al. (2006); $2$ - Lubin et al. (2000); $3$  - Ofek
et al. (2006); $4$ - Eigenbrod
et al. (2006a); $5$ - Rusin et al. (2003); $6$ -Schechter et al. (1998);
 $7$ - Fassnacht \& Cohen
(1998); $8$ - Myers et al. (1999); $9$ - Eigenbrod
et al. (2006b); $10$ - Morokuma et al. (2007); $11$ - Inada et al. (2006);
$12$ - Inada et al. (2007a); $13$ - Burud et al. (2002); $14$ - Ratnatunga et al. (1999); $15$ -
Leh$\acute a$r et al. (1993); $16$ -
Treu \& Koopmans (2003);}
\end{table*}

\section{DATA AND METHOD }
\subsection{The lens galaxy sample}

Our lens sample is primarily drawn from CASTLES (Mu$\mathrm{\tilde n}$oz et
al. 1998) data base\footnote{http://cfa-www.harvard.edu/castles/index.html}. 
While this total sample size of CASTLES is 100 systems,
a statistical analysis requires a sample from a survey that is complete
and has well-characterized, homogeneous selection criteria.
From CASTLES data base, we choose 56 samples discovered (re-discovered)
by SDSS from CASTLES data.
To eliminate biases in terms of the redshift of the lensing galaxy, 
we have selected systems with $0.6<z_{s}<2.2$. 
This source redshift range comes from SDSS selection criterion
of the statistical sample of strongly lensed quasars (Inada et al. 2007b). 

Furthermore we exclude the systems whose lensing galaxies are the spiral
galaxies because the lens model of spiral galaxies is not well established, 
and the fraction of spiral galaxies in the lensing galaxies has not been
decided yet.
Furthermore the contribution of the spiral galaxies on the lensing
probability is expected to be small.
We also exclude the systems whose image separation are large ($>2.0^{''}$) or 
the systems that contain multiple lensing galaxies. 

Using above procedure, our sample contains the lensing systems
whose lens redshift is unknown.
We exclude these lensing systems because we need the lens redshift and
the source redshift information in using this lens-redshift test.        
Lenses at a low redshift are more likely to have a spectroscopically confirmed redshift
than lenses at high redshift so this may bias the sample to systems
where the lens is at an artificially low redshift.
But our aim is to investigate the usefulness of the lens-redshift
test in using the lensing sample from SDSS only.
For this reason, we simply neglect these event at this time.
In the end our lensing sample consists only of the systems with 
single early-type galaxy as lensing object and this contains 13 lensing
systems (see Table 1).

\subsection{The maximum likelihood method}

We use a maximum likelihood estimator in our statistical analysis of lens redshift 
distributions.
In this paper, we follow the method proposed by Kochanek (1992) and 
use the differential optical depth to lensing with respect to 
the angular separation $\Delta\theta$ as the probability density.
For each systems, we calculate the probability density $P(z_{L}|z_{S},\Delta \theta,\ {X})$ 
which is normalized unity, where {X} is the set of galaxy evolution model parameters.
For power-law evolution model, this set is ${X}=(\nu_{n},\ \nu_{v})$, and
for Mitchell's evolution model, ${X}=(p,\ q)$. 
The probability distributions of lens redshifts $z_{L}$ with a given angular separation 
are calculated as follows, 
\begin{equation}
P(z_{L}|z_{S},\ \Delta\theta,\
 {X})=\frac{\frac{d^{2}\tau}{dz_{L}d\Delta\theta}(z_{L},\ z_{S},\
 \Delta\theta,\ X)}{\frac{d\tau}{d\Delta\theta}(z_{S},\ \Delta\theta,\ X)}.
\end{equation}
The likelihood as a function of X is given by,
\begin{equation}\label{eq-like}
L(X) = \prod^{N}_{i=1} P_{i}(z_{L}|z_{S},\ \Delta\theta,\ {X}),
\end{equation}
where N is total system number.

\section{RESULTS} 

We investigate two evolution models by estimating 
the likelihood function (\ref{eq-like}).

First, we obtain constrains on the power-law evolution model
parameters, ($\nu_{n}$, $\nu_{v}$).
We  calculate the $\nu_{n}$-$\nu_{v}$ contours using the SIS lens and the
SIE lens for SDSS DR1 and DR5 data.
The results are shown in Fig. 1.
The maximum L value and the $1\sigma$ limits are
($\nu_{n}=-2.0^{+4.0}_{-3.0}$, $\nu_{v}=0.15^{+0.40}_{-0.35}$) and
($\nu_{n}=-1.0^{+3.5}_{-3.5}$, $\nu_{v}=0.0^{+0.50}_{-0.30}$) using the SIS lens 
for SDSS DR1(top left panel in Fig. 1) data and DR5 data(top right panel in Fig. 1), respectively.
In case of SIE lens model, we have ($\nu_{n}=1.5^{+3.0}_{-3.0}$, $\nu_{v}=-0.15^{+0.20}_{-0.10}$) 
and ($\nu_{n}=2.0^{+4.0}_{-3.0}$, $\nu_{v}=-0.25^{+0.20}_{-0.20}$) for SDSS DR1(bottom left panel in Fig. 1) 
data and DR5 data(bottom right panel in Fig. 1), respectively.

We compare the estimation of parameters used in the power-low evolution model with previous results.
Ofek, Rix \& Maoz (2003) give the 1$\sigma$ limit $\mathrm{d}\log \sigma_{\ast}/\mathrm{dz}=-0.1^{+0.06}_{-0.06}$ for the evolution of the
characteristic velocity dispersion. 
The 1$\sigma$ limit of the characteristic velocity dispersion of the lenses at $z=1$ must
be within 69\%-91\% of the present value.
Meanwhile, Chae \& Mao (2003) obtain limit $\nu_{v}=0.2^{+0.6}_{-1.0}$ (1$\sigma$) without using the lensing rate.
The 1$\sigma$ limit of the characteristic velocity dispersion of the lenses at $z=1$ must be within
57\%-174\% of the present value.
On the other hand, our results are 84\%-104\% (the SIE lens for DR1) and
73\%-97\% (the SIE lens for DR5).
Thus we obtain the constrains on the evolution of the characteristic velocity 
dispersion which are consistent with these two previous studies.

Ofek, Rix \& Maoz (2003) also give the 1$\sigma$ limit $\mathrm{d}\log
\phi_{\ast}/\mathrm{dz}=+0.7^{+1.4}_{-1.2}$ for the number density.
The 1$\sigma$ limit of number density of the lenses at $z=1$ must be
30\% of the present value.
Our results are 35\% (the SIE lens for DR1) and 50\% (the SIE lens for
DR5) and thus are consistent with their result.

\begin{figure*}
\hbox{
\includegraphics[width=75mm]{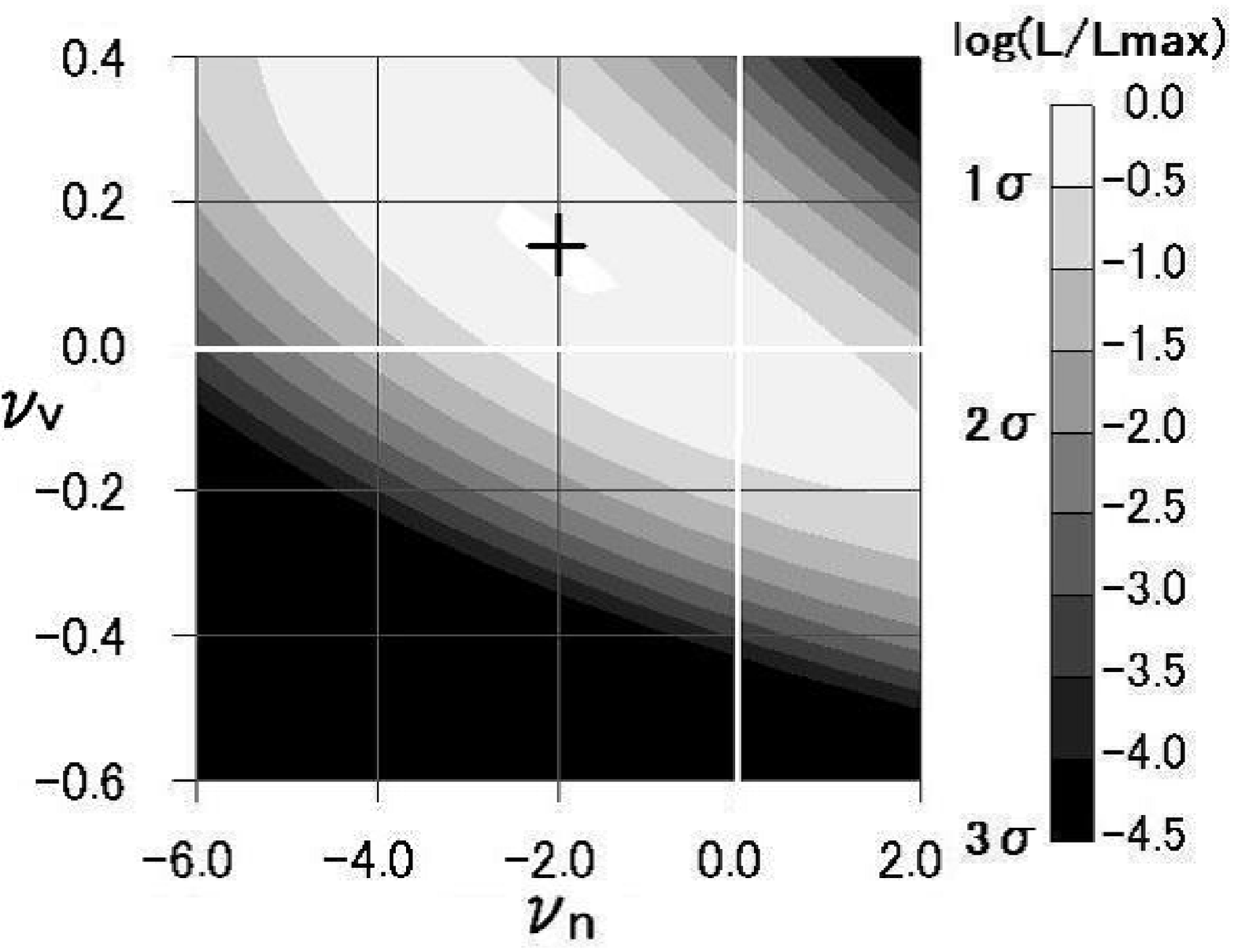}
\hspace{3mm}
\includegraphics[width=75mm]{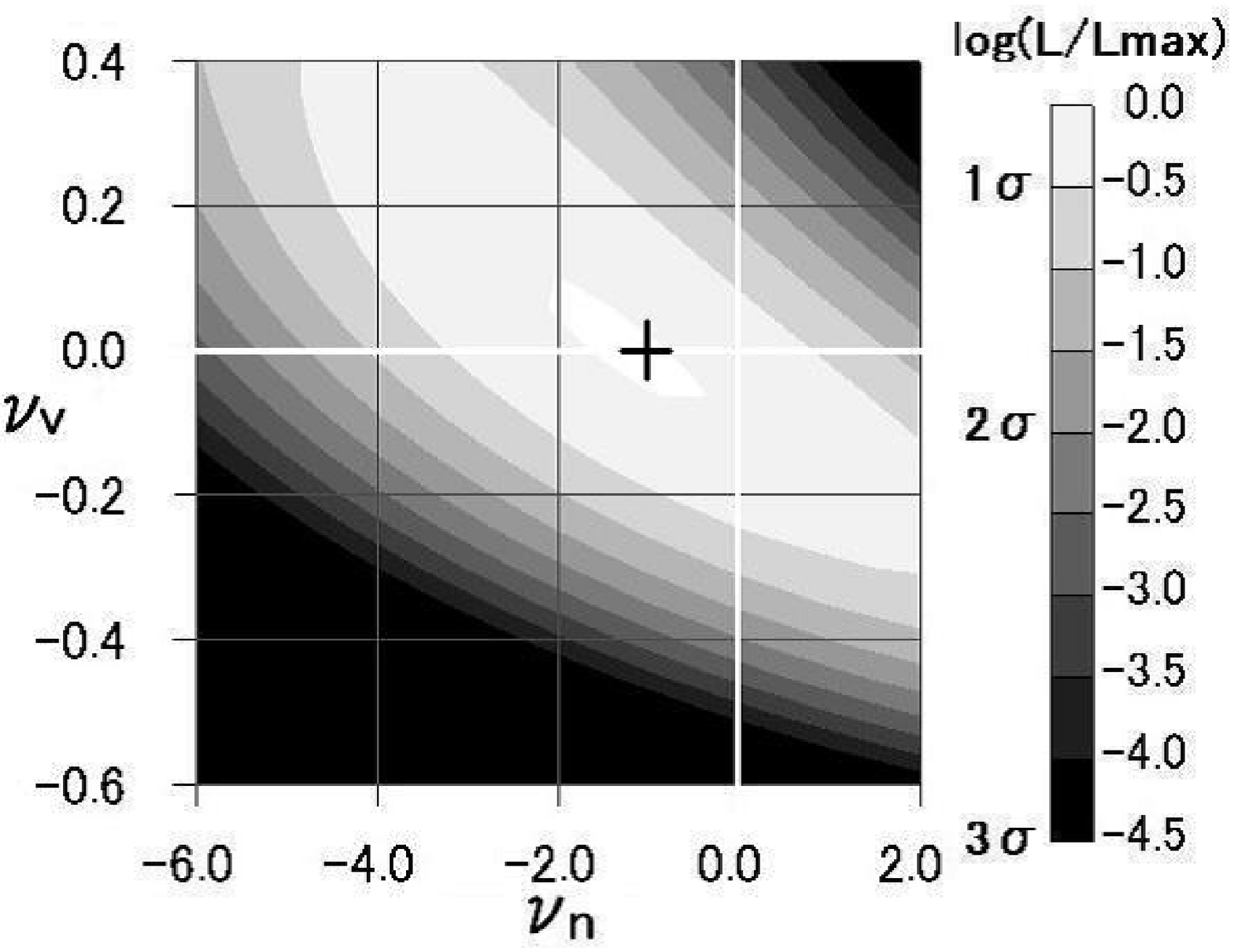}
}	

\hbox{
\includegraphics[width=75mm]{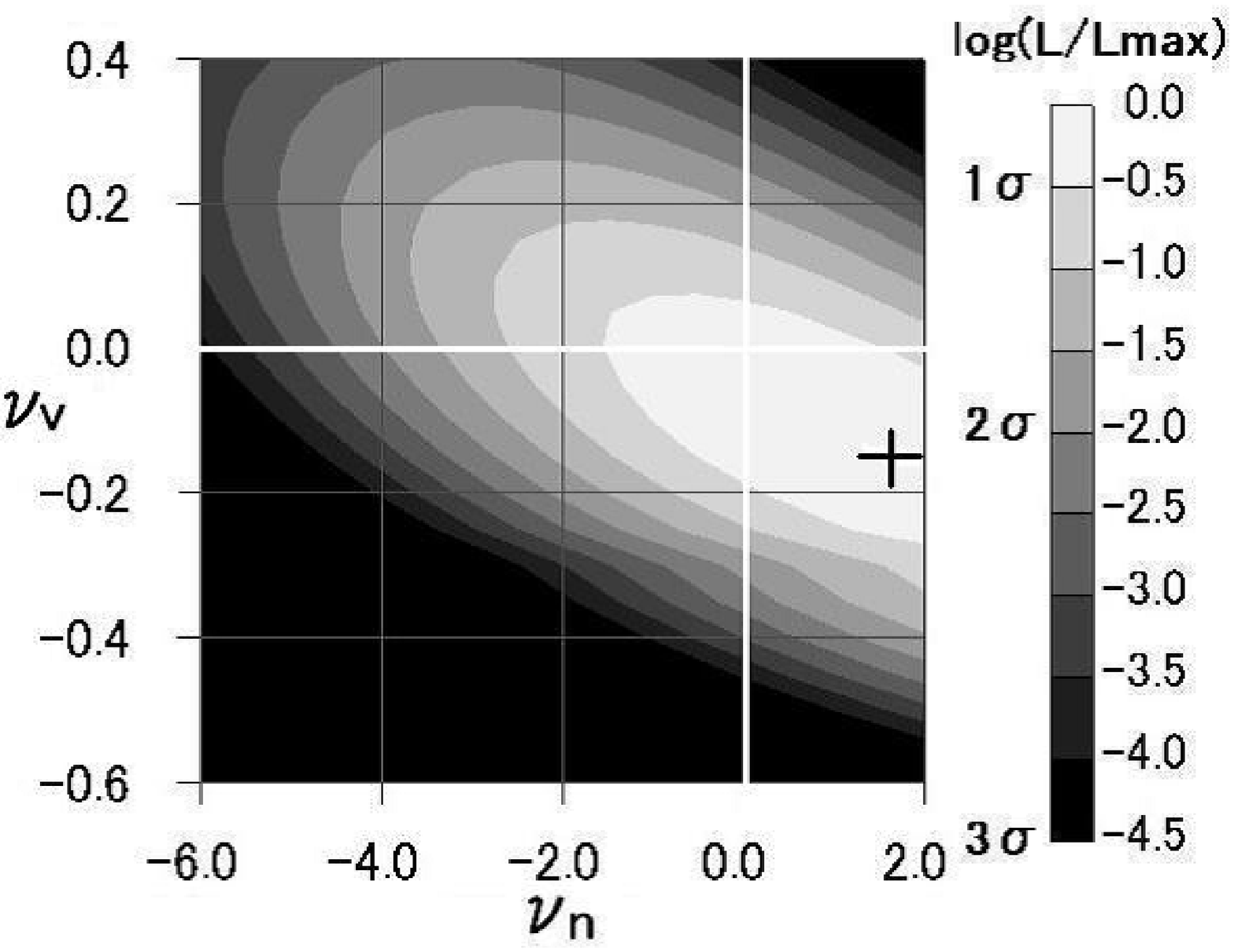}
\hspace{3mm}
\includegraphics[width=75mm]{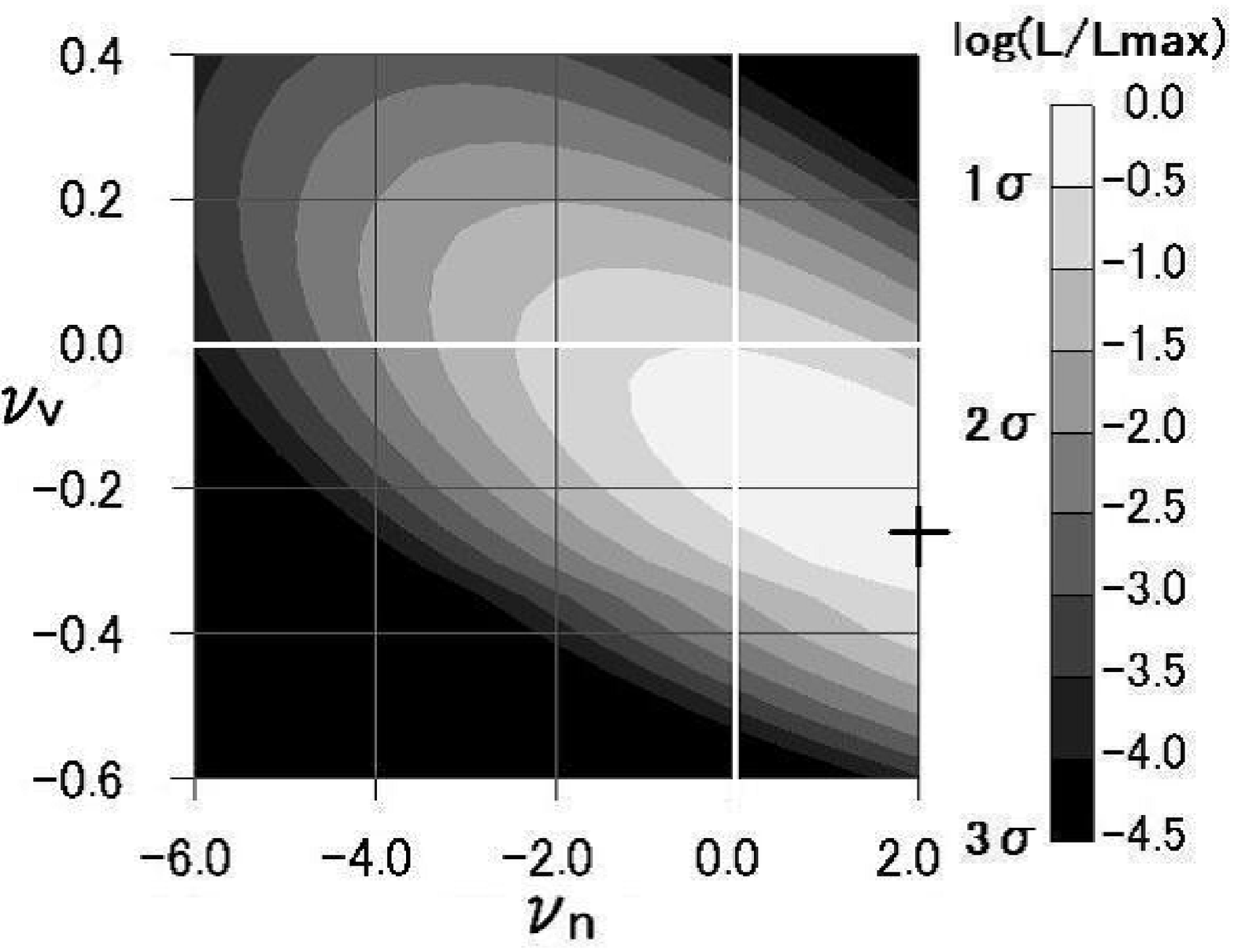}
}	
\vspace{-2mm}

	\caption{The likelihood functions of the number density evolution index
 $\nu_{n}$ and the velocity dispersion evolution index $\nu_{v}$. The
 functions are normalized the corresponding maximum ($L_{max}$). The
 cross represent the maximum likelihood values. Top left panel: We use the SIS lens and the DR1 data as the parameters of
 the velocity function. The maximum value of L is $\nu_{n}=-2.0$, $\nu_{v}=0.15$. Top right panel: We use the SIS lens and the DR5 data as the parameters of
 the velocity function. The maximum value of L is $\nu_{n}=-1.0$, $\nu_{v}=0.0$. Bottom left panel: We use the SIE lens and the DR1 data as the parameters of
 the velocity function. The maximum value of L is $\nu_{n}=1.5$, $\nu_{v}=-0.15$. Bottom right panel: We use the SIE lens and the DR5 data as the parameters of
 the velocity function. The maximum value of L is $\nu_{n}=2.0$, $\nu_{v}=-0.25$.
}
 
\end{figure*}

Next, we calculate the $p$-$q$ contours using the SIS lens and the
SIE lens for SDSS DR1 and DR5 data.
We limit our investigation to $p<5$ and $q<3$ because the value of
$\phi(\sigma;\ z)$ become unphysical beyond this range.
The reason is as follows. 
We consider redshift evolution of the number density of lensing  galaxies 
by multiplying current velocity function, (\ref{eq:velocity function}), 
by the velocity function ratio, $\phi(\sigma;z)/\phi(\sigma;z=0)$ (see (2.4.2)) 
just like Mitchell et al. (2005).
If we try to calculate the ratio in the case of $p>5$, the resulting ratios
become negative. Moreover the parameter $\alpha$ in the velocity
function becomes negative in the case of $q>3$.

Our results of the likelihood functions of the parameters, ($p$, $q$) , are
in Fig. 2.
In Fig. 2, we obtain maximum values in L at ($p=5.0$, $q=0.1$) for the
SIS lens and ($p=4.5$, $q=0.1$) for the SIE lens 
which are largely different from the estimation from the simulation 
($p=0.3$, $q=0.75$ the white-dots in Fig. 2).

The maximum value, p and q are at the very corner of the parameter
space so we expect that our best fit values of p and q are the maximum of
the likelihood 
within our physically meaningful limits and it is probably not a true
maximum of the likelihood.
But we think it still can be said  that the values for p and q used
by Mitchell et al. are inconsistent with the presently available
observational data.

Furthermore we compare the power-law evolution model and the model given
by Mitchell et al. ($p=0.3$, $q=0.75$). 
We use our best fit parameter value of the power-law evolution model, $\nu_{v}$, in the
case of SIE lens, ($\nu_{v}=-0.15$ for DR1 and $\nu_{v}=-0.25$ for DR5)
and calculate the ratio of the velocity function at $z=0.3$ and $z=1.0$ to that at $z=0.0$.
The error of the parameter $\nu_{n}$ is large, so we use only estimated
$\nu_{v}$ and set $\nu_{n}=0$.
Similarly, we use the Mitchell et al.'s evolution model($p=0.3$, $q=0.75$)and calculate the ratio of the velocity function.
This result is shown in Fig.3. 
As a result, we find that the evolution model by the Mitchell et al.
 has possibility to underestimate the number density of galaxy whose velocity dispersion is large.

\begin{figure*}
\hbox{
\includegraphics[width=75mm]{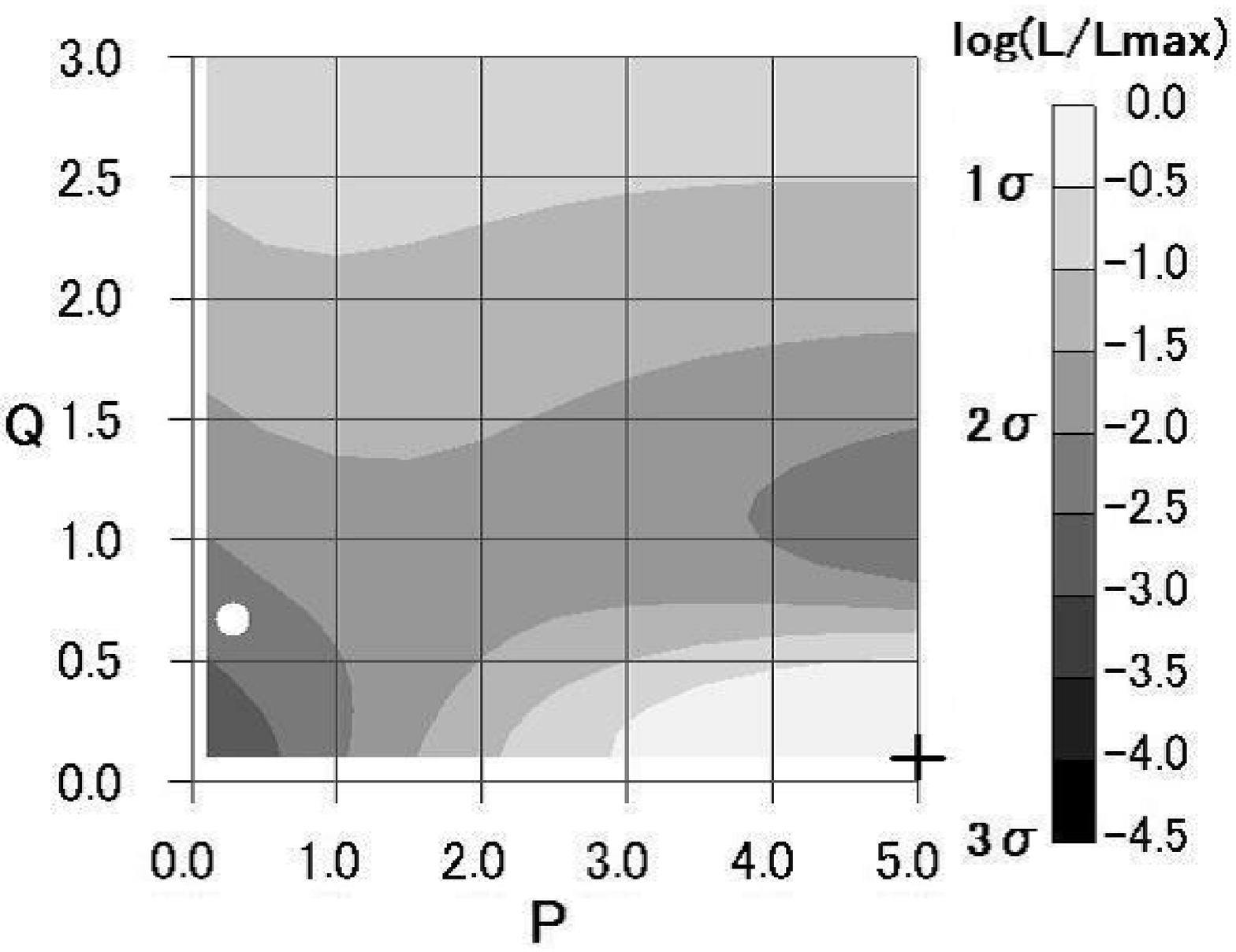}
\hspace{3mm}
\includegraphics[width=75mm]{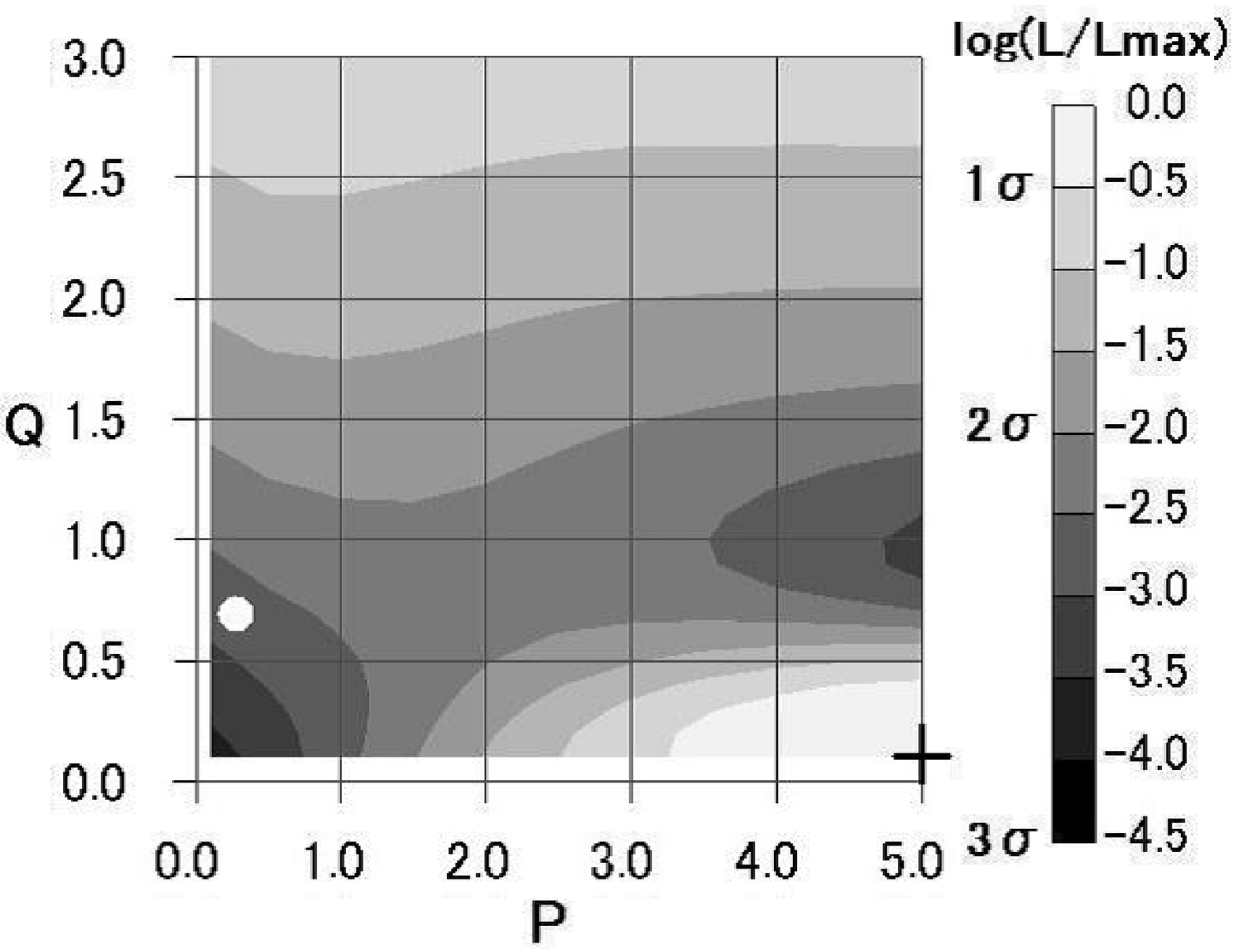}
}
\hbox{
\includegraphics[width=75mm]{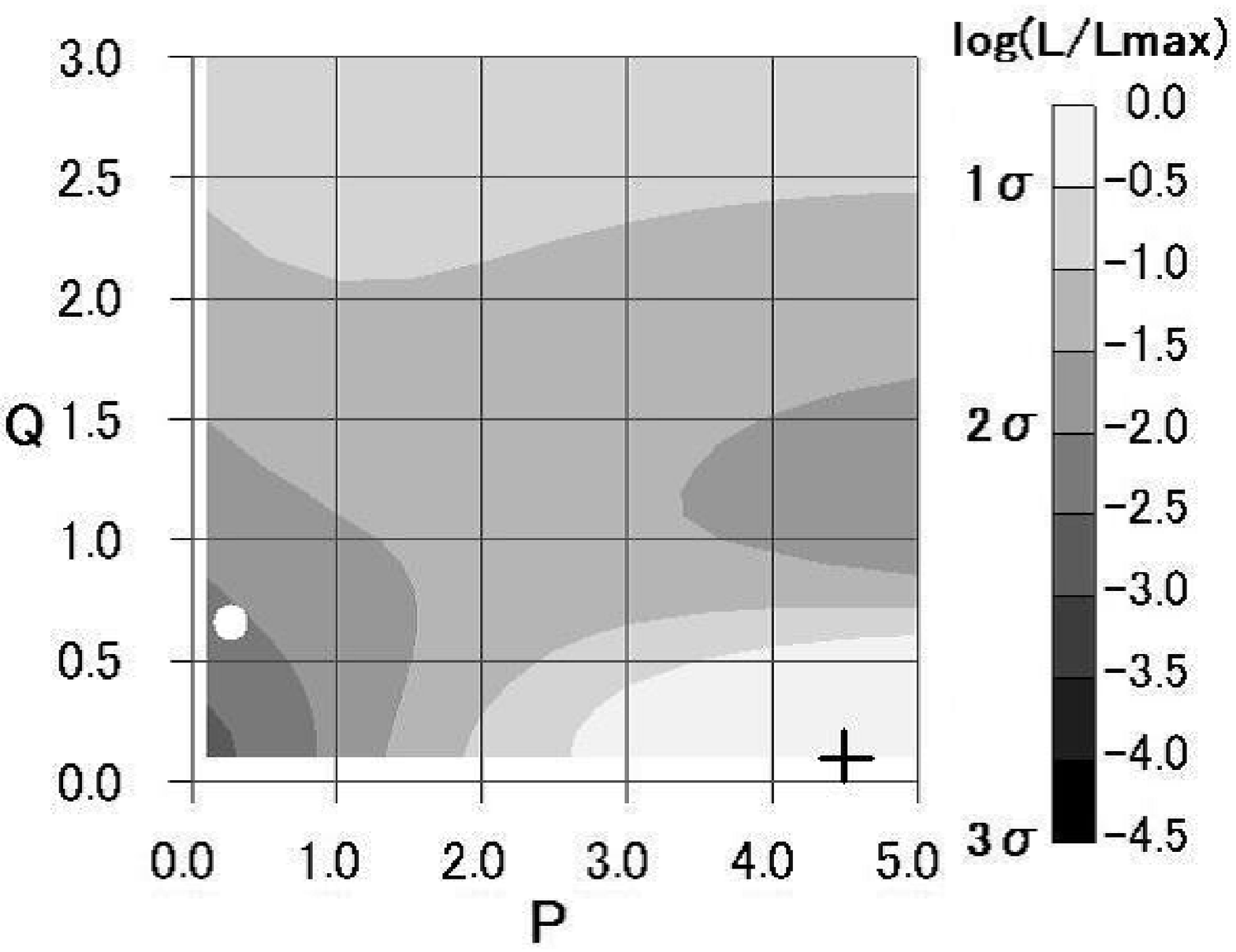}
\includegraphics[width=75mm]{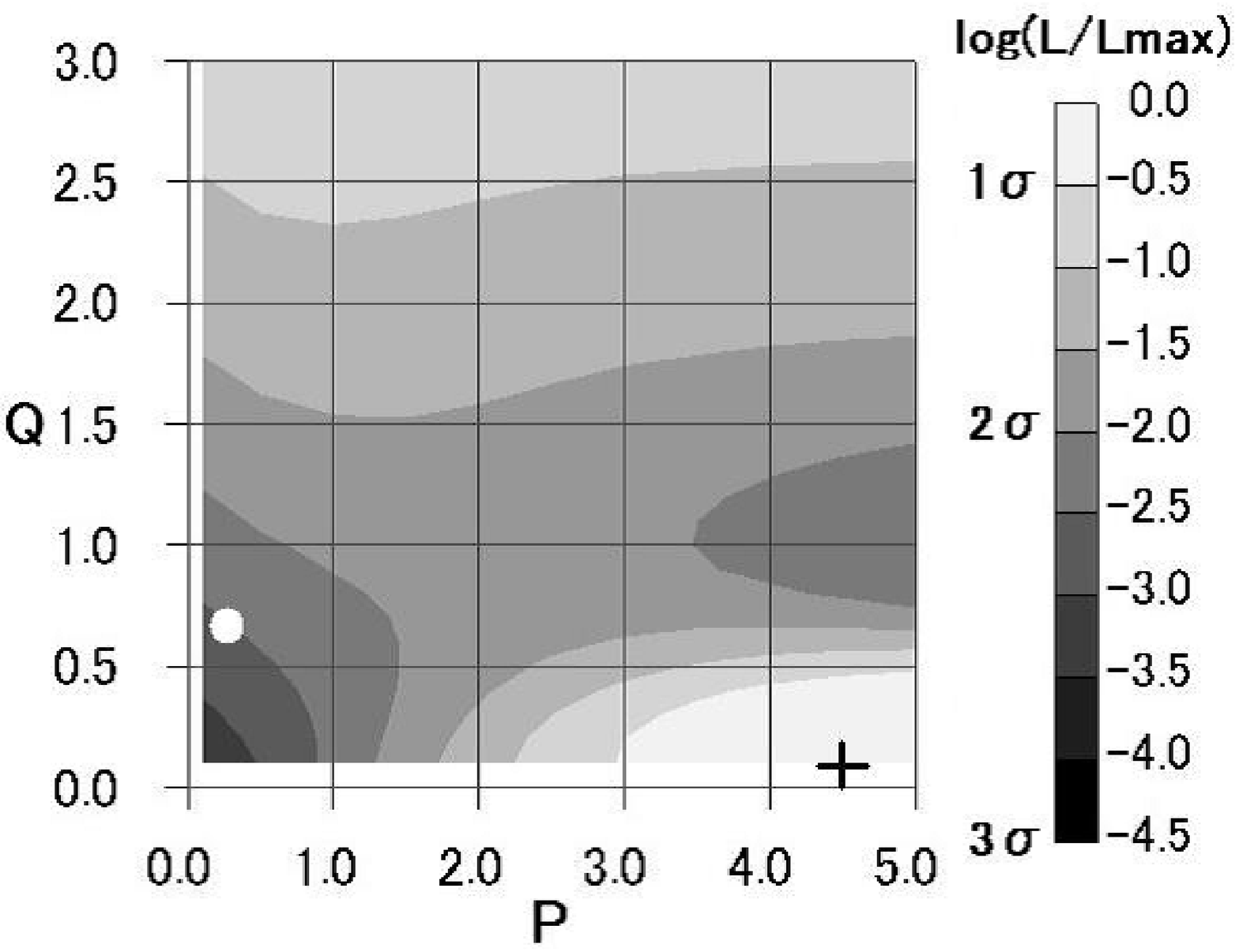}
}	
\vspace{-2mm}

  \caption{The likelihood functions of the parameters, (p, q) used in Mitchell's
 evolution model. The
 functions are normalized the corresponding maximum ($L_{max}$).
Top left panel: We use the SIS lens and the DR1 data as the parameters of
 the velocity function. The maximum value of L is $P=5.0$,
 $Q=0.1$.
Top right panel: We use the SIS lens and the DR5 data as the parameters of
 the velocity function. The maximum value of L is $P=5.0$,
 $Q=0.1$.
Bottom left panel: We use the SIE lens and the DR1 data as the parameters of
 the velocity function. The maximum value of L is $P=4.5$,
 $Q=0.1$.
Bottom right panel: We use the SIE lens and the DR5 data as the parameters of
 the velocity function. The maximum value of L is $P=4.5$,
 $Q=0.1$.
The white points represent estimation from
 simulation.
}

\end{figure*}

\begin{figure*}
\hbox{
\includegraphics[width=55mm,angle=270]{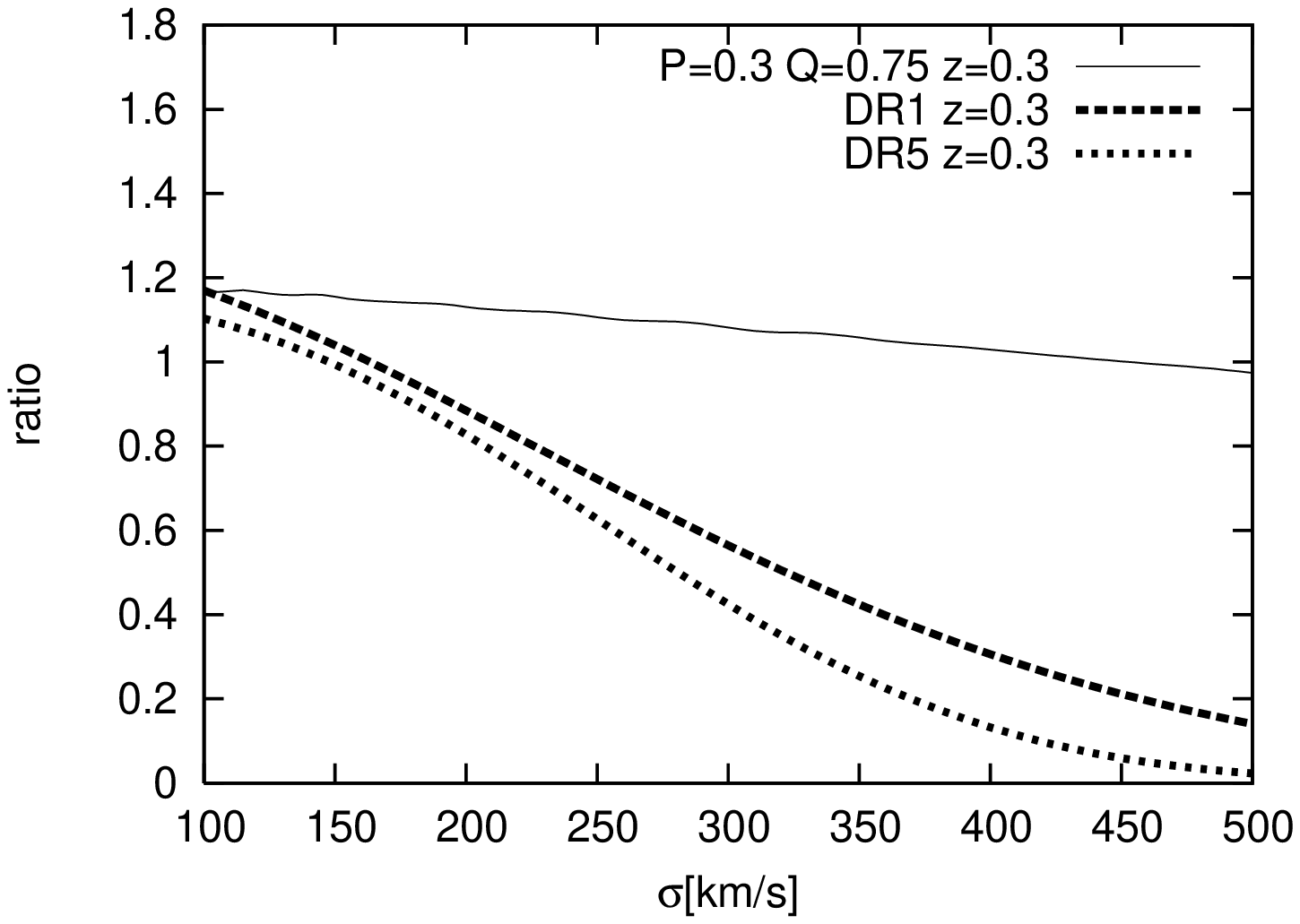}
\hspace{2mm}
\includegraphics[width=55mm,angle=270]{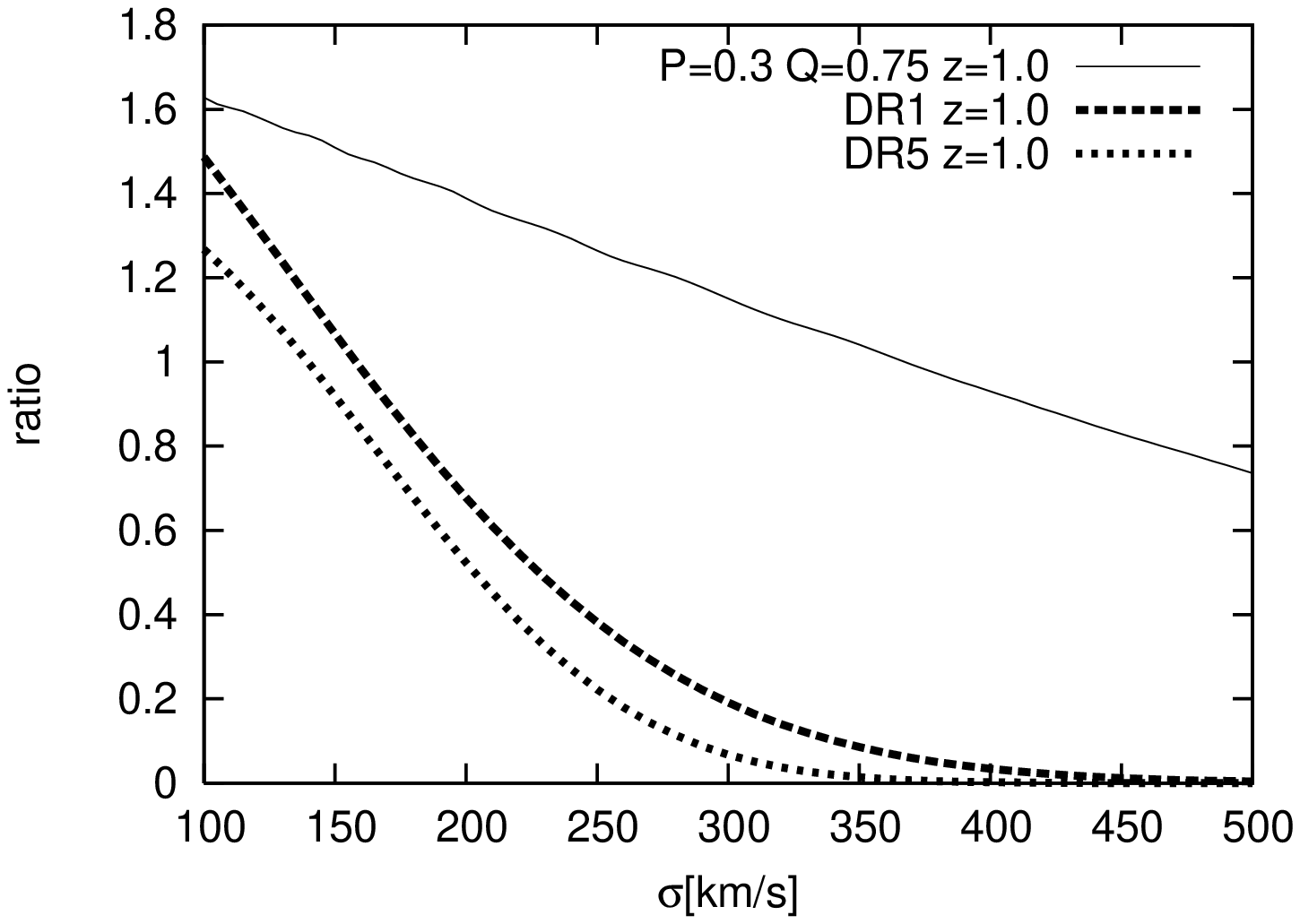}
}

\caption{
Ratio of the velocity function at
 $z=0.3$ and $z=1.0$ to that at $z=0$. The set, ($p=0.3$, $q=0.75$) is estimated by
 numerical simulated values.
And DR1 and DR5 in the figure show the ratio of velocity function 
when we use the
 best fit values estimated by using SIE lens model,
 ($\nu_{n}=0.0$, $\nu_{v}=-0.15$) and ($\nu_{n}=0.0$, $\nu_{v}=-0.25$),
 respectively. (see figure 1 and section 4)}

\end{figure*}

\section{CONCLUSIONS AND DISCUSSION}

We employed the `lens-redshift test` proposed by Kochanek (1992) and 
constrain the parameters of two evolution models, the power-low evolution model 
and the evolution model proposed by Mitchell et al. (2005).
We use the SDSS lens sample that is well-defined.
It turns out that one of the parameters of the power-law evolution
model, $\nu_{v}$ is consistent with that given 
by Ofek, Rix \& Maoz (2003) and Chae \& Mao (2003) and the other parameter $\nu_{n}$ is 
consistent with the result by Ofek, Rix \& Maoz (2003).

On the other hand, the most likely values for the parameters used 
in the evolution model by Mitchell et al. (2005) are inconsistent with
the predicted values from the presently available observational data.
The inconsistency may be mainly caused by ignoring the total lensing rate in
the present analysis. Although Chae \& Mao (2003) have shown that the
evolution model can be restricted more strictly by including the information
of lensing rate, we have not
included the effect because the total rate depends on the magnification
bias which can not be calculated at the moment with very good accuracy
in a consistent way with our SDSS sample.
Furthermore we find that there is a possibility to underestimate the
number density of galaxy which have large velocity dispersion 
by using the parameters in Mitchell et al.'s evolution model.

The SQLS is still ongoing and the
number of the strong lensing events is steadily increasing. Thus if we are
able to calculate accurate
magnification bias based on SDSS sample, our method
will be very useful to constrain the evolution models of the number density
of the lensing galaxies and thus the CDM based structure formation scenario.

\section*{Acknowledgements}
We thank Masamune Oguri for helpful discussions and the
anonymous referees for helpful comments that improved our conclusion.


\begin{thebibliography}{99}
\bibitem[\protect\citeauthoryear{Asada et al. 2003}{}]{b2}Asada H.,
		Hamana T., Kasai M., 2003, A\&A, 397, 825
\bibitem[\protect\citeauthoryear{Augusto et al. 2001}{}]{b22}Augusto P. et al., 2001, MNRAS, 326, 1007
\bibitem[\protect\citeauthoryear{Browne et al. 2003}{}]{b24}Browne I. W. A., et al., MNRAS, 2003, 341, 13
\bibitem[\protect\citeauthoryear{Bryan \& Norman 1998}{}]{b10}Bryan
		G. L., \& Norman M. L., 1998, ApJ, 495, 80
\bibitem[\protect\citeauthoryear{Burud et al. 2002}{}]{b48}Burud I., et
		al., 2002, A\&A, 391, 481
\bibitem[\protect\citeauthoryear{Capelo \& Natarajan 2007}{}]{b17}
		Capelo P. R., \& Natarajan P., 2007, in preparation (astro-ph/07053042v1)
\bibitem[\protect\citeauthoryear{Chae 2007}{}]{b7}Chae K. -H., 2007,
		ApJ, 658, L71
\bibitem[\protect\citeauthoryear{Chae \& Mao 2003}{}]{b6}Chae K. -H.,
		\& Mao S., 2003, ApJ, 599, L61
\bibitem[\protect\citeauthoryear{Chae, Mao \& Kang 2006}{}]{b16} Chae K. -H., Mao S., \& Kang X., 2006, MNRAS, 373, 1369
\bibitem[\protect\citeauthoryear{Choi \& Vogeley 2007}{}]{b5} Choi Y. -Y., Park, C., \& Vogeley M. S., 2007, ApJ, in press
		(astro-ph/0611607)
\bibitem[\protect\citeauthoryear{Eigenbrod et
			al. 2006a}{}]{b35}Eigenbrod A., Courbin F., Dye S., Meylan G., Sluse D., Vuissoz
		C., Magain P., 2006, A\&A, 451, 747
\bibitem[\protect\citeauthoryear{Eigenbrod et al. 2006b}{}]{b43}Eigenbrod
		A., Courbin F., Meylan G., Vuissoz
		C., Magain P., 2006, A\&A, 451, 759
\bibitem[\protect\citeauthoryear{Fassnacht \& Cohen
		1998}{}]{b40}Fassnacht C. D., Cohen J. G., 1998, AJ, 115,
			377
\bibitem[\protect\citeauthoryear{FFKT
		1992}{}]{b40}Fukugita. M.,  Futamase. T., Kasai. M., \& Turner. E. L., 
        1992, ApJ, 393, 3			
\bibitem[\protect\citeauthoryear{Inada et al. 2006}{}]{b31}Inada N., et
		al., 2006, AJ, 131, 1934
\bibitem[\protect\citeauthoryear{Inada et al. 2007a}{}]{b27}Inada N., et
		al., AJ, 2007, 133, 206
\bibitem[\protect\citeauthoryear{Inada et al. 2007b}{}]{b28}Inada N., et
		al., astro-ph, 0708.0828v1
\bibitem[\protect\citeauthoryear{Kochanek 1992}{}]{b29}Kochanek C. S.,
		1992, ApJ, 384, 1
\bibitem[\protect\citeauthoryear{Lacey \& Cole 1994}{}]{b12} Lacey C.,
		\& Cole S., 1994, MNRAS, 271, 676
\bibitem[\protect\citeauthoryear{Lehar et al. 1993}{}]{b50}Lehar
		J., Langston G. I., Silber A., Lawrence C. R., Burke
		B. F., 1993, AJ, 105, 847
\bibitem[\protect\citeauthoryear{Lubin 2000}{}]{b33}Lubin L. M., Fassnacht
		C. D., Readhead A. C. S., Blandford R. D., Kundi$\it
		\acute c$ T., 2000, AJ, 119, 451
\bibitem[\protect\citeauthoryear{Mao 1991}{}]{b13} Mao S., 1991, ApJ,
		380, 9
\bibitem[\protect\citeauthoryear{Mao \& Kochanek 1994}{}]{b14} Mao S.,
		\& Kochanek C. S., 1994, MNRAS, 268, 569
\bibitem[\protect\citeauthoryear{Maoz \& Rix 1993}{}]{b30}Maoz D., \&
		Rix H -W., 1993, ApJ, 416, 425
\bibitem[\protect\citeauthoryear{Mitchell et al. 2005}{}]{b4}Mitchell,
		J. L., Keeton C. R., Frieman J. A., Sheth R. K., 2005, ApJ, 622, 81
\bibitem[\protect\citeauthoryear{Morokuma et al. 2007}{}]{b44}Morokuma et
		al., 2007, AJ, 133, 214
\bibitem[\protect\citeauthoryear{Mu$\mathrm{\tilde n}$oz et.al. 1998}{}]{b11}Mu$\mathrm{\tilde n}$oz J. A.,
		Falco E. E., Kochanek C. S., Leh$\acute a$r J., McLeod
		B. A., Impey C. D., Rix H. -W., Peng C. Y., 1998,
		Ap\&SS, 263, 51
\bibitem[\protect\citeauthoryear{Myers et al. 1999}{}]{b42}Myers S. T.,
		et al., 1999, AJ, 117, 2565
\bibitem[\protect\citeauthoryear{Myers et al. 2003}{}]{b23}Myers S. T.,
		et al., 2003, MNRAS, 341, 1
\bibitem[\protect\citeauthoryear{Newman \& Davis 2000}{}]{b9}Newman J. A., \& Davis, M., 2000, ApJ, 534, L11
\bibitem[\protect\citeauthoryear{Ofek, Rix \& Maoz 2003}{}]{b15} Ofek
		E. O., Rix H. -W., \& Maoz D., 2003, MNRAS, 343, 639
\bibitem[\protect\citeauthoryear{Ofek et al. 2006}{}]{b34}Ofek E. O., Maoz
		D., Rix H. -W., Kochanek C. S., Falco E. E., 2006, ApJ,
		641, 70
\bibitem[\protect\citeauthoryear{Oguri et al. 2006}{}]{b26}Oguri M., et
		al., AJ, 2006, 132, 999
\bibitem[\protect\citeauthoryear{Oguri et al. 2007}{}]{b27}Oguri M., et
		al., astro-ph/0708.0825v1
\bibitem[\protect\citeauthoryear{Patnaik et al. 1993}{}]{b20}Patnaik
		A. R., Browne I. W. A., King L. J., Muxlow T. W. B.,
		Walsh D., \& Wilkinson P. N., 1993, MNRAS, 261, 435
\bibitem[\protect\citeauthoryear{Pantaik et al. 1992}{}]{b21}Pantaik A. R., Browne I. W. A., Walsh D., Chaffee F. H., \&
		Foltz C. B., 1992, MNRAS, 259, P1
\bibitem[\protect\citeauthoryear{Ratnatunga et
		   al. 1999}{}]{b49}Ratnatunga K. U., Griffiths
			   R. E., Ostrander E. J., 1999, AJ, 117, 2010
\bibitem[\protect\citeauthoryear{Riess et al. 1998}{}]{b18} Riess
		A. G. et. al., 1998, AJ, 116, 1009
\bibitem[\protect\citeauthoryear{Rusin et al. 2003}{}]{b36}Rusin D., Kochanek
		C. S., Keeton C. R., 2003,ApJ, 595, 29
\bibitem[\protect\citeauthoryear{Schechter et al. 1998}{}]{b37}Schechter
		P.L., Gregg M. D., Becker R. H., Helfand D. J., White
		R.L., 1998, AJ, 115, 1371
\bibitem[\protect\citeauthoryear{Sheth et al. 2003}{}]{b3} Sheth
		R. K. et al., 2003, ApJ, 594, 225
\bibitem[\protect\citeauthoryear{Sheth \& Tormen 1999}{}]{b8} Sheth
		R. K.,\& Tormen G., 1999, MNRAS, 308, 119
\bibitem[\protect\citeauthoryear{Spergel et al. 2007}{}]{b52}Spergel
		D.F., et al., 2007, ApJS, 170, 377S
\bibitem[\protect\citeauthoryear{Tonry 1998}{}]{b47}Tonry J. L.,
		1998, AJ, 115, 1
\bibitem[\protect\citeauthoryear{Treu \& Koopmans 2003}{}]{b51}Treu
		T., Koopmans L. V. E., 2003, MNRAS, 343, L29
\bibitem[\protect\citeauthoryear{Turner, Ostriker \& Gott 1984}{}]{b1}
		Turner E. L., Ostriker J P., \& Gott, J. R.,
		I\hspace{-.1em}I\hspace{-.1em}I., 1984, ApJ, 284, 1
\bibitem[\protect\citeauthoryear{York et al. 2000}{}]{b25}York D. G.,
		et al., AJ, 2000, 120, 1579





\end{thebibliography}
\end{document}